# Radiation dose simulation during laser-plasma proton acceleration experiments and method to increase the measurement resolution of the proton energy spectrum

M. Ganciu[1*,] A. Chirosca[1,4], A. Groza[1], E. Stancu[2,5], O. Stoican[3],

D.B. Dreghici[1,5], B. Butoi[1], C. Ticos[6], B. Cramariuc[7]

[1] National Institute for Laser, Plasma and Radiation Physics (INFLPR), Low Temp. Plasma Dept., Atomistilor Str. No. 409, 077125 Magurele, Ilfov County, Romania
[2] National Institute for Laser, Plasma and Radiation Physics, STARDOOR Dept., Atomistilor Str. No. 409, 077125 Magurele, Romania
[3] National Institute for Laser, Plasma and Radiation Physics (INFLPR), Plasma Physics and Nuclear Fusion Laboratory, Atomistilor Str. No. 409, 077125 Magurele, Ilfov County, Romania
[4] Faculty of Physics, University of Bucharest, Nuclear Phys. Dept. Magurele, Romania
[5] Faculty of Physics, University of Bucharest, Doctoral School in Physics, Magurele, Romania
[6] National Institute for Laser, Plasma and Radiation Physics, (INFLPR), Accelerator Dept., Atomistilor Str. No. 409, 077125 Magurele, Romania
[7] IT Center for Science and Technology, (CITIST), Bucharest, Romania
* Correspondence: mihai.ganciu@inflpr.ro;

## Abstract

The paper discusses some 3D simulations to compute the ionizing radiation dose during laser-plasma experiments leading to the generation of accelerated protons and electrons. Also, we suggest a new method to increase the measurement resolution of the proton energy spectrum. Monte-Carlo simulations of the radiation doses map around the laser-foil interaction point are performed using Geant4 General Particle Source code and the particular geometry chosen for the experimental setup. We obtain the map of the radiation dose distribution for high-power laser - thin solid target experiments, considering a cubic geometry of the interaction chamber. The computed radiation dose distribution shows a good agreement with various, previously obtained experimental results, and could be a step towards simulating the radiation environment inside of a spacecraft. To characterize the laser-plasma accelerated protons, we introduce a new method to enhance the measurement resolution of the proton energy spectrum by employing a stack of thin solid detectors, preferably CR-39. Each detector is thinner than the Bragg peak region on the Bragg curve that characterizes the loss of kinetic energy as a function of the particle penetration depth in the detector material. The relevance of this method for space radiation characterization and cybersecurity insurance is highlighted.

**Keywords**: laser-accelerated proton beams; CR-39 detectors; ionizing radiation; high-power laser; Geant4 General Particle Source code

## 1. Introduction

Space weather represents a branch of space physics and aeronomy (heliophysics), that deals with the investigation of the continuously changing environment conditions characteristic to the Solar system. It includes the study of phenomena that occur in the magnetosphere, ionosphere, thermosphere, and exosphere.

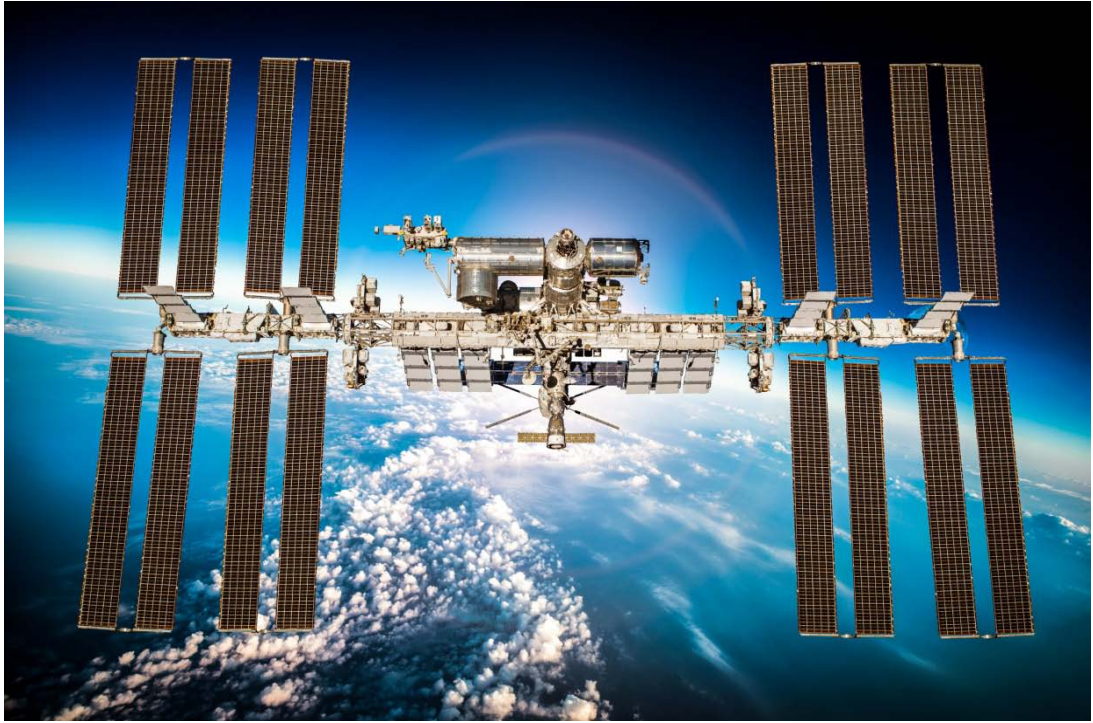

Fig. 1: The International Space Station during its travel around the Earth is strongly affected
by the intense ionizing radiation during solar storms (Credit Shutterstock - 240333836)

Space weather is determined by the solar wind and the interplanetary magnetic field carried by the solar wind hot plasma (Zell, 2017). A variety of physical phenomena are associated with space weather, including geomagnetic storms, or constant remodelling of the Van Allen radiation belts caused by interplanetary shock waves - an outburst of highly energetic particles carried by the solar wind which can literally blast away the outer radiation belt and then split its remains into two distinct rings (Baker et al., 2013). Other phenomena associated with space weather are ionospheric disturbances and scintillation of satellite-to-ground radio signals and long-range radar signals (Cannon, 2013), auroras and geomagnetically induced currents at Earth's surface. Coronal mass ejections and the associated shock waves can also result in major perturbation of space weather, as they can compress the magnetosphere and trigger geomagnetic storms. Solar energetic particles accelerated by coronal mass ejections or solar flares can trigger solar particle events, a critical factor concerning the impact of space weather on humans, as they can inflict damage on electronics onboard spacecraft, endanger the lives of astronauts, or increase radiation hazards for high-altitude, high-latitude air flights.

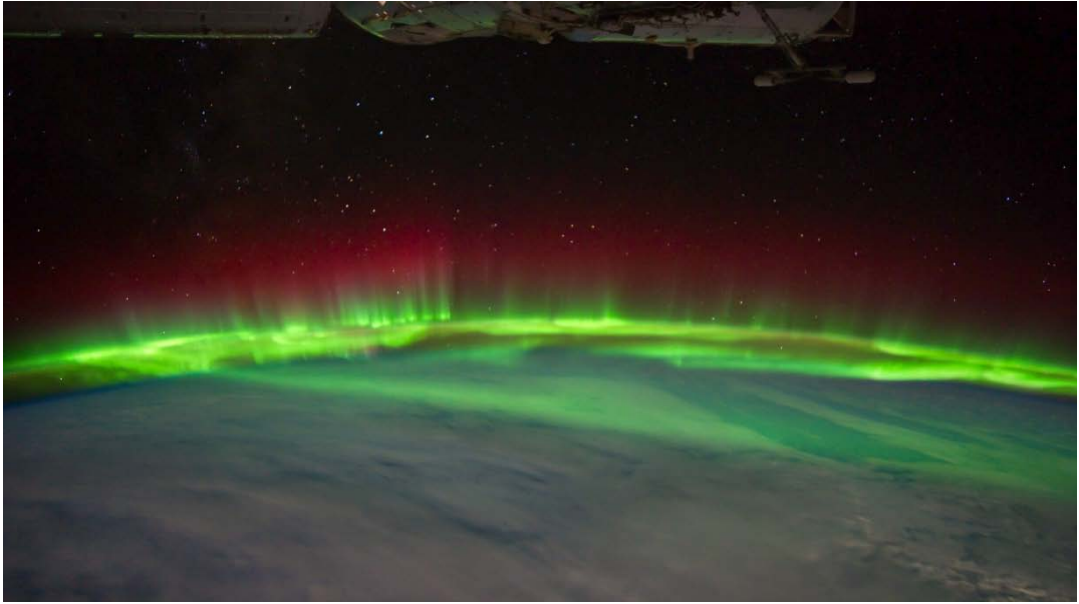

Fig. 2: Aurora borealis from the
International Space Station (Credit Shutterstock - 240333836)

Spacecraft charging, the accumulation of electrostatic charge on non-conducting materials at the spacecraft surface due to low energy particles is one of the predominant space weather effects on an orbiting spacecraft (Delzanno, Borovsky, Thomsen, Moulton, & Macdonald, 2015). The cumulative damage contributions of the different particle species are evaluated in terms of dose equivalent. The space environment exhibits a low dose rate of ~ (10-6 -10-4) Gy/s. As the duration of space missions spreads over the years or even decades, the total dose can be very large as a result of the successive accumulation process. During the active period (lifetime) of a near-Earth space mission, total ionizing doses of 103 Gy are a common figure. One of the most unpleasant results consists of the radiation effects on solid-state microelectronics. They can be divided into two categories: 1. Cumulative effects; 2. Single event effects (SEE). Cumulative effects cause gradual changes in the operational parameters, while SEE causes abrupt changes or transient behavior in circuits. For charged particles, the amount of ionizing energy deposited in a material is related to the stopping power or the linear energy transfer function in that material, commonly expressed in units of MeV/cm2. Assessing radiation sensitivity levels for electronic components is a fundamental issue in space-systems radiation hardness assurance programs. Preliminary results obtained in INFLPR show that simple electronic circuits introduced in a 6 MeV electron beam are rendered inoperable by a dose of (100 - 300) Gy delivered within a few seconds (Ticoş et al, 2019). One important issue lies in developing redundant software algorithms and artificial intelligence approaches for space equipment, to ensure cybersecurity during intensive radiation exposure.

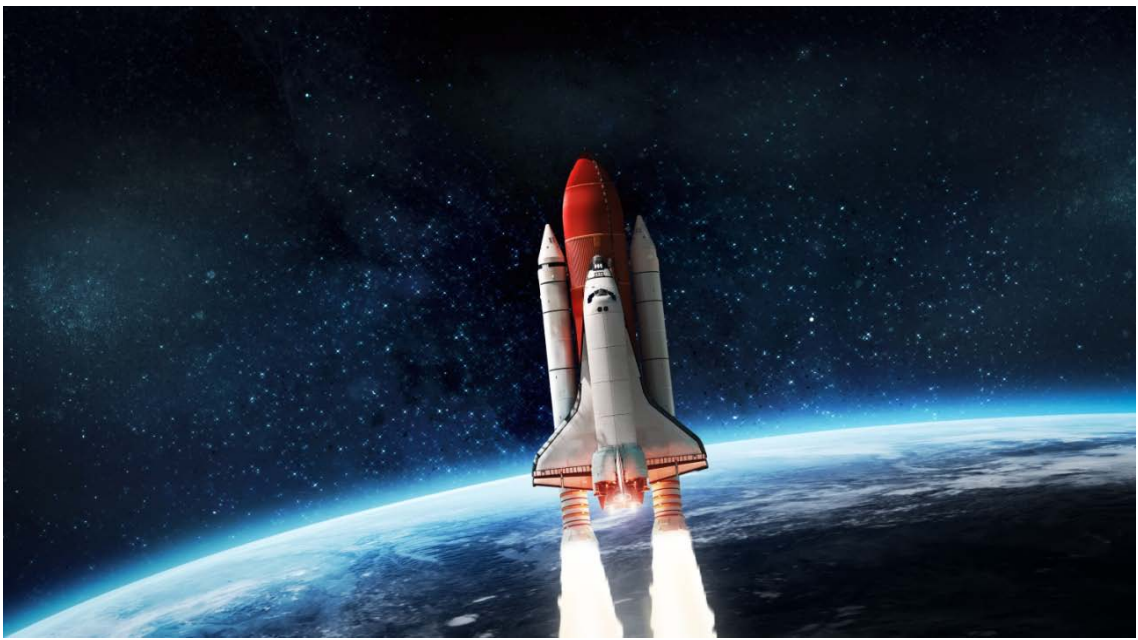

Fig. 3: Space shuttle in its travel in the aggressive radiological environment
specific to the cosmic space (Credit Shutterstock - 240333836)

Another harmful effect of ionizing radiation is the exposure of the human body to space radiation
(Zell, 2017; Delzanno, 2015).

Conditions similar to the radiation environment characteristic to space stations and spacecraft can be
achieved using high power/high-intensity lasers (Koenigstein, Karger, Pretzler, Rosenzweig, &
Hidding 2012; Ganciu, M., et al., 2015). High-intensity lasers are used in current research to study
matter under extreme conditions and for laser-driven particle acceleration. A high-intensity laser pulse
creates plasma at the target surface, and then interacts with it, producing "hot" electrons. As a result
of this interaction, some electrons are accelerated up to energies of tens of MeV  (Macchi, Borghesi,
& Passoni, 2013; Tampo, et al., 2010). The "hot" electrons and laser-accelerated electrons interact
with the target chamber walls, generating X-ray photons by bremsstrahlung. The mixed field of
photons and electrons might create a radiation hazard associated with such laser-matter interaction
experiments (Koenigstein, et al., 2012). Therefore, different ionizing particle types and doses can be
obtained in high power laser - thin solid target experiments. For example, protons with energies up to
tens of MeV and mixed fields of electrons and photons with doses of tens of mGy/shot can be
generated, depending on the target thickness and material (Koenigstein, et al., 2012). Appropriate
experimental conditions can be achieved at several high power laser facilities worldwide (Danson,
Hillier, Hopps, & Neely, 2015; Asavei et al., 2016).

The paper presents simulations of integrated doses produced by the mixed field of electrons and X-ray
photons generated during high power laser-thin solid target interaction experiments, in which laser-accelerated protons with energies up to about 14 MeV (Ganciu et al. 2019 a; Groza et al., 2019) were demonstrated by using the CETAL high power laser facility (CETAL, 2019).

## 2. Radiation dose simulation

The interaction of high-power lasers with thin solid targets generates protons, electrons, X-ray radiation, and neutrons (by using a converter, e.g., a Be foil) that spans a wide energy range (Fuchs et al., 2006; Hidding et al., 2017). Laser accelerated electrons that interact with the walls of the target chamber produce X-ray radiation by bremsstrahlung. In our simulation, we are using the PW-CETAL laser parameters reported in (CETAL, 2019; Giubega, 2018; Groza et al., 2019).

The experimental conditions enable one to reconstruct the radiation field within the interaction chamber assimilated with a cube, which allows us to characterize specific radiation conditions (such as particle flow and dose) for any point in space. Dose reconstruction has been achieved by simulating the radiation transport using the Geant4 V.10.5 framework. To obtain a realistic simulation, two-volume sources are considered.

The first source would be the pre-plasma obtained due to the interaction of the PW laser with the target surface. Hot, relativistic electrons are generated with a temperature of 2.8 MeV. The source is considered to exhibit a uniform angular distribution and a Gaussian energy distribution. The source location is chosen in front of the target (Tampo et al., 2010; Xiao et al., 2018). The second source, considered as an accelerated proton source, is located on the rear side of the target.

A more detailed simulation was performed showing the radiation effects for a small tissue sample (5 x 5 x 3 cm3 volume) shielding with a 2 mm Al shield with a 1 mm W coating on the interface between the Al and the tissue. The proton distribution was generated using Target Normal Sheet Acceleration (Fuchs et al., 2006). An Al target of 10 µm thickness was considered, in experimental conditions similar to those described in (Jeong et al., 2017; Volpe et al., 2019). The distance between the source and the multi-layered detector was fixed at 1.5 cm leading to the spatial dose distribution presented in Fig. 4.

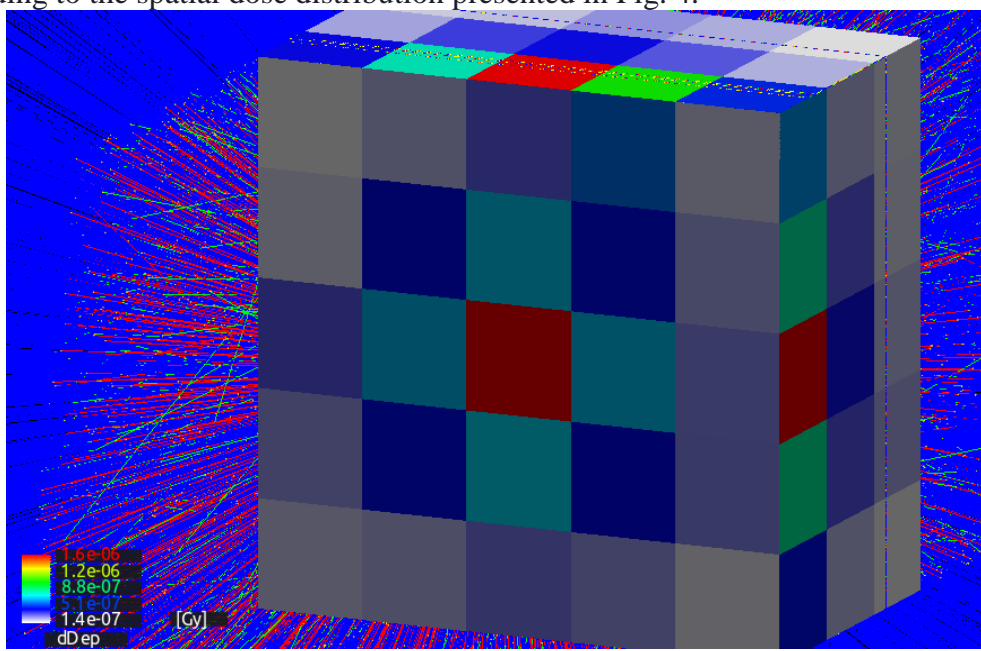

Fig. 4: Total volumetric dose distribution estimate for the

tissue detector (scaled x 1E5)

The radiation produced by the interaction of the High Energy Laser with the thin target was described by the means of two sources implemented using the Geant4 General Particle Source module, with a 10:1 ratio (the main particles are the protons, our simulation generates one electron primary for every 10 protons generated primaries). One of the sources generates relativistic electrons in the point of interaction and the other one produces a proton beam with the specific TNSA spectra. The protons that originate from the source propagate through the environment and undergo nuclear reactions when they interact with the experimental setup within the simulation region, generating secondary radiation such as photons and electrons. The electron, proton, and photon dose depth profile is presented in Fig. 5. Here, the dose due to the gamma radiation is significantly lower (7E-7 Gy) than the one produced by protons (up to 0.35 Gy/pulse) and electrons (0.1 Gy/pulse). As radiation propagates among the Z-axis, the dose depth profile was performed in this direction and the results present the doses integrated among the other axes.

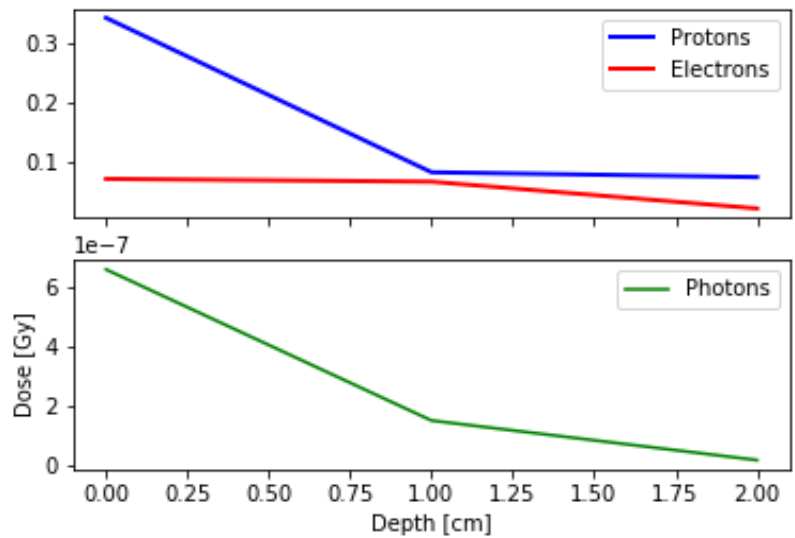

Fig. 5: Partial dose depth profile for each of the radiation type within our detector

A 3D depth profile is presented in Fig. 6. The tight focus can be observed to give the maximum dose in the middle of our detection geometry and presents the way the dose is distributed among the depth (Z direction) and height (Y direction) of our setup.

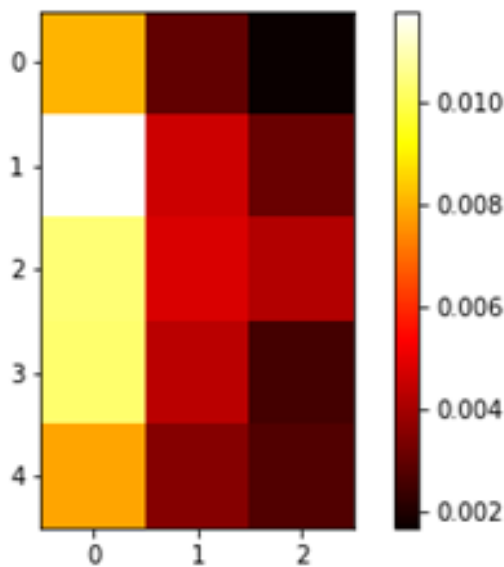

Fig. 6: 3D depth profile of total dose for our simulated detector

This type of simulations can provide a solid dose distribution assessment for high power laser setups allowing for a specific, consistent radiation field to be applied to electronics, equipment, or even organic material. Thus, this type of setup can estimate the space-like conditions and establish the setup requirements (fields, shielding, and placement) of the desired system to obtain the specific radiation field needed to simulate the experimental conditions.

## 3. Method to increase the proton energy spectrum resolution

A patent application was submitted (Ganciu , 2019 b) that refers to a method used to determine the energy distribution of a proton beam. The method, while not being limited to, can be applied to diagnose protons that are (i) generated by the interaction between a high power laser beam and plasma, (ii) present in cosmic radiation, or (iii) generated by the medical equipment employed for proton therapy.

One of the methods used to detect protons consists of using solid media, specifically designed for this purpose and produced as foils that are exposed to a flux of protons. Currently, the most prevalent substance used to this end is *poly allyl diglycol carbonate* (C12H18O7) commercially known as CR-39. The protons crossing these detectors deteriorate the structure of the material and leave physical traces in the form of sensitized areas of energetic electrons that emerge as an outcome of the interaction between the proton and the foil material. They represent electrons that are "extracted" from the atoms, with a specific energy that dissipates in the neighboring regions. These electrons sensitize the physical structure of the material by breaking some of the chemical bonds and thus rendering the material easier to corrode in the specific etching process. Different etching agents are used, such as NaOH or KOH. The etching agent acts upon the foil (detector) surface, as the corrosion rate is higher for the sensitized areas where micro-craters appear. These features are visualized and counted using a microscope. In (Jeong et al., 2017) a method is described to determine the energetic spectrum of protons at normal incidence on a CR-39 detector, based on counting the micro-craters generated by the protons. For a 1 mm thick CR-39 plate, an upper limit of the maximum energy is reported at 7.5 MeV. In our case, the analysis of the micro-craters on the back of the plate indicates the presence of protons with energies of about 10 MeV (Groza et al, 2019).

Owing to a variation of the proton stopping position by tenths of micrometers, caused by statistical fluctuations of the proton energy transfer towards the material they cross, micro-craters with different dimensions occur for a 20 mm etching, even if they result from protons of equal energy. This represents proof that protons can exhibit energies larger than 10 MeV. Protons that cross into the second CR-39 detector leave identifiable traces only on the back surface of the first detector. Nevertheless, these protons can be identified by analyzing the traces left on the front surface of the second CR-39 detector, according to the method suggested in (Groza et al., 2019). The paper also shows the issues associated with determining the energetic spectrum for protons with energies between (7.5 - 10) MeV. To obtain reproducible results the etching conditions such as time of application concentration, and temperature, should be identical. By analysing the shape and dimensions of the traces, one can determine the energy and the direction of motion for incident protons.

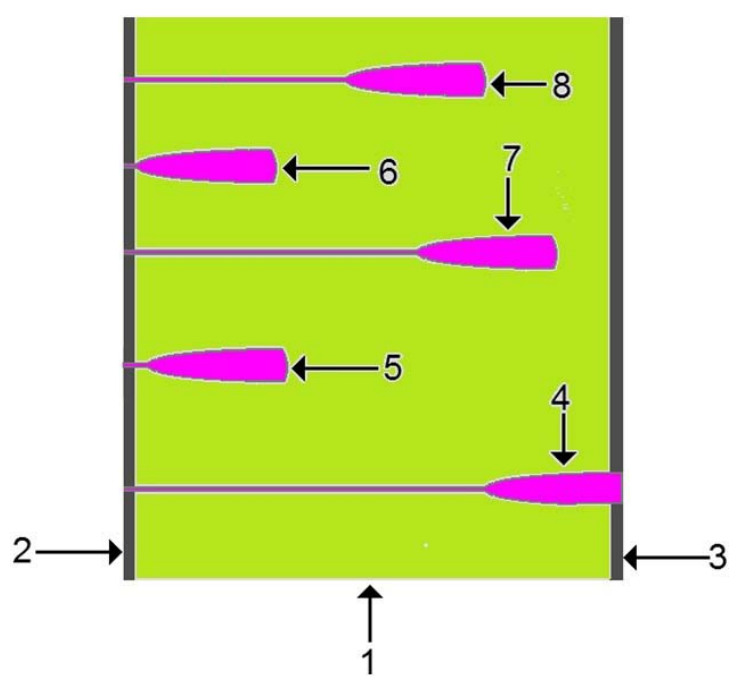

Fig. 7: Representation of the affected zone (4,5,6,7,8) following the stopping of the protons with various energies, in the CR-39 (1). The etched layers are also represented (2,3)

According to (Sinanian, et al., 2011), by using 1500 µm thick CR-39 nuclear track detectors, it is possible to determine the proton energy distribution in the range (0.92 - 9.28) MeV.

Accurate determination of the proton energetic spectrum, and particularly of the maximum energy acquired after the laser-pulse solid target interaction, enables one to estimate the electron temperature as shown in (Tampo et al., 2010). This issue is very important for developing applications of interest for Radiation Hardness Assurance, by simulating the aggressive radiation environment from the outer space using laser-plasma accelerators (Koenigstein et al., 2012). The method supplies information about the characteristics of individual protons. Nevertheless, the method described in (Jeong et al., 2017) also presents some disadvantages such as:

- The determination of incident proton characteristics is done only for particles whose energy lies within a specified value range.

- To visualize and measure the trace dimensions at the foil, the etching process must be performed in several steps, at well-defined intervals of time. After each step, the foil surface is optically examined then the etching process is resumed, which finally leads to an increase in the processing time.

The method proposed to determine the energetic spectrum of protons is based on using a set of solid detectors, preferably CR-39. Each detector is thinner than the Bragg peak on the Bragg curve that characterizes the loss of kinetic energy as a function of the particle track in the detector material. The Bragg peak is an intense peak whose position is a clear indication of the particle location where the particle loses the largest part of its kinetic energy. For protons, the Bragg peak occurs right before stopping. The method we devised circumvents some of the disadvantages associated with previous methods, and brings in the following advantages:

- It enhances the resolution used to determine the proton energy and extends the interval of energy values for which the method is applicable. These characteristics are useful for proton beams with energy spread over a wide range of values.

- The etching process is simultaneously performed for all solid detectors. Consequently, etching is performed faster, under identical conditions for all solid CR-39 detectors. Each detector is etched only once.

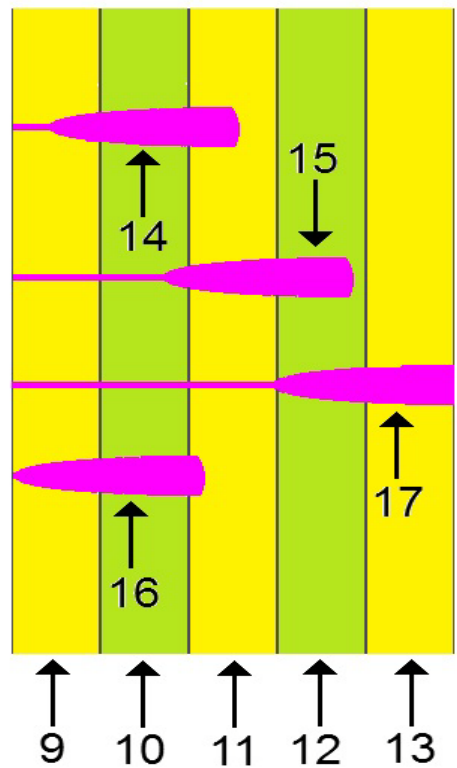

Fig. 8: Affected zones (14,15,16,17) following the stopping of the protons with various energies, in a stack of CR-39 foils (9,10,11,12,13)

The thickness of the Bragg peak region is defined as the full width at half maximum (FWHM) of the energy loss curve, which describes the specific energy deposition dE/dx of a heavy charged particle as a function of the penetration depth. For proton energies in the (3 - 14.8) MeV range, it is about 0.2 mm. Thus, the thickness of the CR-39 detector should not exceed 0.2 mm.

As an example of applying this method to determine the proton energy spectrum, a stack of 10 CR-39 200 µm thick detectors is used. The stack is wrapped within a 10 µm thick aluminium foil for protection. A preliminary calibration procedure is mandatory, using protons with well-known energy. This could be for instance achieved by using the accelerators located at (IFIN-HH, 2019) that can generate protons of 2, 3 and 5 MeV. Moreover, an accelerator can be used that delivers protons with the energy of ~ 12 MeV. By passing a monoenergetic proton beam through materials with different widths, one can obtain an exponential proton energy spectrum. Passive detectors such as CR-39 are time-independent, thus if the proton flux is low enough for avoiding nonlinear effects they behave as if all protons arrived at the same time. In such conditions, calibration is performed by exposing the CR-39 stack to a flux of protons at normal incidence, and the fluence is tuned such that the micro-craters do not overlap. The detectors are etched using a classical procedure (Groza et al., 2019), then optical microscopy is used to analyze the traces left on both front and back faces of the solid detector. The number of micro-craters on each face of the detector is counted and analyzed, to emphasize experimental patterns. We need to bring into consideration that protons lose ~1 MeV when crossing the 10 mm Al foil. Measurement errors are determined by considering previously performed calibrations.

## 4. Discussions and Conclusion

For the estimated maximum energy of the protons of approximately 14 MeV, the obtained simulation results suggest that high power laser - thin solid target experiments can be used as a test environment for the electronics and the software supposedly to be deployed in spacecraft and space station habitats.

The simulated dose is due to mixed X-ray photons and electron radiation field. A mesh of the experimental set-up geometry is generated and used as input for computing the radiation doses around the interaction point, by using the Geant4 General Particle Source code. The resulting radiation dose map shows good accordance with the previous experimental results.

We extend our assumption and claim that the code can also be used for other types of radiation sources to simulate radiation inside and near-space stations, at least by considering a superposition of radiation fields, if nonlinear effects can be neglected.
Even if the space radiation background is pseudo-continuous, while the one generated in laser-targeted interaction experiments comes in short pulses, the key advantage of laser-plasma interaction is that it can generate simultaneously electrons, protons, and X-rays.

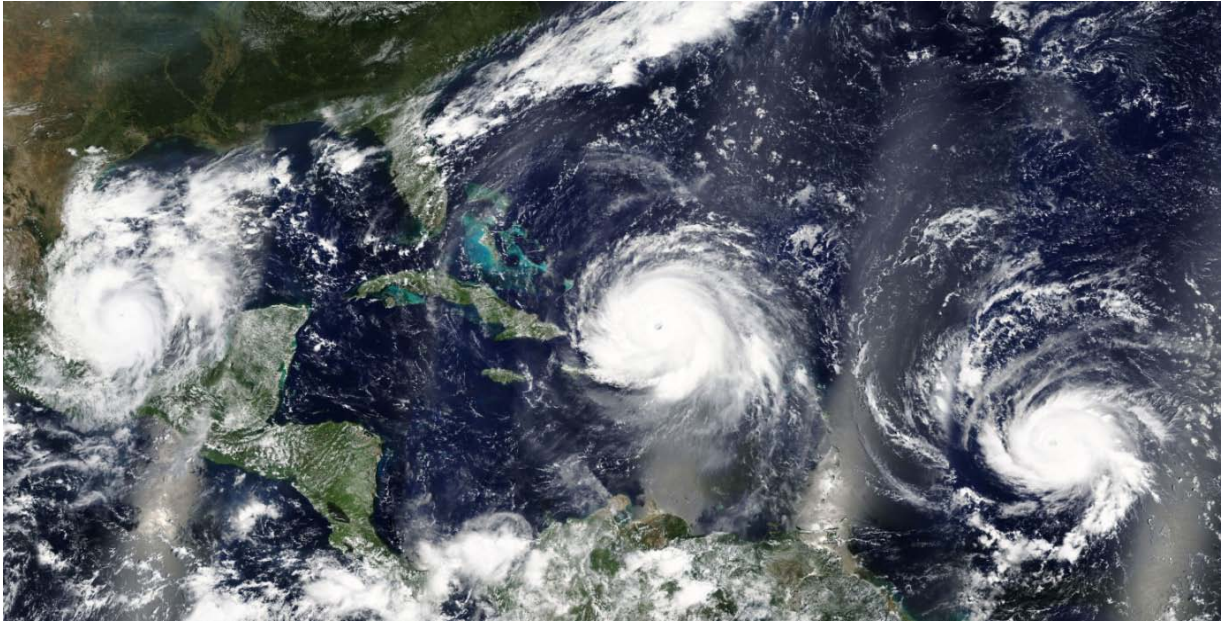

Fig. 9: Safe weather monitoring is provided by satellites that are resistant to the aggressive radiological cosmic environment  (Credit Shutterstock - 240333836)

Also, new shielding materials and designs of current interest could be tested.

To increase the resolution of the proton energy spectrum determination, a new method is developed, characterized by using plastic foil stacks (preferably CR-39), but not limited to. The thickness of each foil should be less than the one of the major-specific energy deposition region of a proton, i.e., around the Bragg peak.

This method for determining the proton energy spectrum is characterized by using stacks of plastic foils for which the etching is performed on each face, to a depth of less than 20 μm, avoiding damage of the foil's mechanical characteristics.

The number of protons stopped in each foil is assessed by analyzing the number of pits corresponding to the protons stopped on the front side of the foil. Only those tracks are considered that have a diameter larger than a certain value, which is characteristic for tracks induced by protons at the limit of stopping when passing through a given foil.

Particularly, the plastic film stack could be used to determine the proton energy spectrum when testing the effect of proton fluxes on electronic components. This plastic foil stack could also be used for the characterization of laser-accelerated protons and for medical applications (e.g., proton therapy), as well as for monitoring certain environments which are subject to ionizing radiation, such as space stations, ships and shuttles, especially during solar storms or the passage of a space vehicle through the radiation belts of some planets in the solar system.

The stacks of plastic films can be calibrated with proton beams of known energy distribution, generated in classical or modern proton accelerators. A mechanical device could be used to rapidly insert various filters between the calibration proton source and the foil stack. This way, the energy of protons reaching the stack is reduced, in a known and controllable way.

The proton energy spectrum is determined based on the analysis of protons stopped in each foil. Due to the low foil thickness, a well-defined energy can be assigned to each of them, based on its position within the stack. The total number of protons incident on the stack front surface is obtained by summing up the number of protons stopped in each of the foils. The resolution of the assessed proton energy spectrum can be achieved by analyzing the track diameters on the foils.

**Conflicts of Interest:** The authors declare no conflict of interest. The funders had no role in the design of the study; in the collection, analyses, or interpretation of data; in the writing of the manuscript, or in the decision to publish the results.

**Acknowledgments:** We acknowledge Dr. George Nemes, ELI-NP Romania, and ASTiGMAT[TM], USA, and Dr. Șerban Udrea, GSI-Darmstadt, Germany for critical reading of the manuscript.

**Funding:** This work has been funded by European Space Agency within the ESA contract No.4000121912/17/NL/CBi by Romanian National Authority for Scientific Research and Innovation, contract No. 3N/2018 (Project PN 18 13 01 03) and by Romanian Space Agency, contract No. 53/19. 11. 2013 (Competence Center: Laser-Plasma Acceleration of Particles for Radiation Hardness Testing - LEOPARD)
**References**

Allison J. Amako, K., Apostolakis, J., Arcee, P., Asai, M., Asog,T.,... &Yoshida, H., (2019) *Recent developments in Geant4*, Nucl. Instrum. Meth. Phys. Res. A: Accelerators, Spectrometers, Detectors and Associated Equipment **835**, p. 186 - 225, https://doi.org/10.1016/j.nima.2016.06.125 (2016) .

Asavei T., Tomut, M., Bobeica, M. Aogaki S. , Cernaianu M. O., Ganciu,... & Zamfir, N. V., (2016), *Materials in extreme environments for energy, accelerators and space applications at ELI-NP*, Romanian Reports in Physics, **68**, Supplement, P. S275–S347 (2016)

Baker D. N., Kanekal S. G., Hoxie V. C., Henderson M. G., Li X., Spence H. E., Elkington S. R., .... & Claudepierre S. G.(2013), *A long-lived relativistic electron storage ring embedded in Earth's outer Van Allen belt,* Science **340** (6129):186–190

Cannon, P. (2013). *Extreme space weather: impacts on engineered systems and infrastructures.* In Royal Academy of Engineering. Retrieved from
http://www.raeng.org.uk/news/publications/list/reports/Space_Weather_Full_Report_Final.PDF
CETAL, (2019), Center for Advanced Laser Technologies, Ultra-intense Lasers Laboratory; http://cetal.inflpr.ro/newsite/cetal-pw

Danson, C., Hillier, D., Hopps, N., & Neely, D. (2015). Petawatt class lasers worldwide. High Power Laser Science and Engineering, 3, https://doi.org/10.1017/hpl.2014.52

Delzanno, G. L., Borovsky, J. E., Thomsen, M. F., Moulton, J. D., & Macdonald, E. A. (2015). *Future beam experiments in the magnetosphere with plasma contactors: How do we get the charge off the spacecraft?* Journal of Geophysical Research: Space Physics, 120(5), 3647–3664. https://doi.org/10.1002/2014JA020608

Fuchs, J., Antici, P., D'Humières, E., Lefebvre, E., Borghesi, M., Brambrink, E., … Audebert, P. (2006). *Laser-driven proton scaling laws and new paths towards energy increase.* Nature Physics, 2(1), 48–54. https://doi.org/10.1038/nphys199

Ganciu, M. Piso, M.-I. Stoican, O. Mihalcea, B Diplasu, C. Marghitu, O.,... & I. Morjan, (2015) Application System and method for testing components, circuits and complex systems using synchronized and pulsed fluxes consisting of laser accelerated particles, Patent Application, WO201503061 A1, WO2015030619A4 (2015)

Ganciu, M.; Groza, A.; Cramariuc, O.; Mihalcea, B.; Serbanescu, M.; Stancu, E.,...; & Cramariuc,B.(2019a); *Hardware and software methods for radiation resistance rising of the critical infrastructures*. Rom. Cyber Secur. J. **1**, 3–13.

Ganciu,, M., Groza, A., Stoican, O.,Chirosca, A., Dreghici, D., Stancu, E., (2019b), OSIM Patent Applications, A/00337/05-06-19

Giubega, G. (2018). Proton acceleration in ultra-intense laser interaction with solid targets at CETAL-PW laser, Poster presentation, Workshop CETAL**,** http://cetal.inflpr.ro/newsite/workshop

Groza, A. Serbanescu, M. Butoi, B. Stancu, E. Straticiuc, Burducea M. I., …& Ganciu, M.(2019), *Advances in Spectral Distribution Assessment of Laser Accelerated Protons using Multilayer CR-39 Detectors,* Appl. Sci. 9, 2052  (2019);  https://doi.org/10.3390/app9102052

Hidding, B., Karger, O., Königstein, T., Pretzler, G., Manahan, G. G., McKenna, P., & Daly, E. (2017) *Laser-plasma-based Space Radiation Reproduction in the Laboratory*. Scientific Reports, 7 (February), 1–6. https://doi.org/10.1038/srep42354

IFIN-HH (2019). Horia Hulubei National Institute for R&D in Physics and Nuclear Engineering, http://www.nipne.ro/ (last seen Aug 2019).

Jeong, T. W.. Singh, P. K , Scullion, C., Ahmed, H. Hadjisolomou, Jeon, P. C.,... & Ter-Avetisyan, S., CR-39 track detector for multi-MeV ion spectrocopy, **Scientific Reports**, 7:2152 (2017)

Koenigstein T. Karger O., Pretzler G., Rosenzweig J. B., & Hidding B., (2012) *Design considerations for the use of laser-plasma accelerators for advanced space radiation studies,* Journal of Plasma Physics 78 (4), 383 - 391 https://doi.org/10.1017/s0022377812000153

Macchi, A., Borghesi, M., & Passoni, M. (2013). Ion acceleration by superintense laser-plasma interaction. Reviews of Modern Physics, 85(2), 751–793.

Najafi, M., Geraily, G., Shirazi, A. Esfahani, M., & Teimouri, J. (2017), Analysis of Gafchromic EBT3 film calibration irradiated with gamma rays from different systems: Gamma Knife and Cobalt-60 unit, Medical Dosimetry

Sinanian, N. Rosenberg, M. J. Manuel, M. McDuffee, S. C. Casey, D. T. Zylstra, A. B.,... &. Petrasso, R. D (2011)*, The response of CR-39 nuclear track detector to 1-9 MeV protons,* **Rev. Sci. Instrum**., 82, 103303

Tampo, M., Awano, S., Bolton, P. R., Kondo, K., Mima, K., Mori, Y., … &Kodama, R. (2010). Correlation between laser accelerated MeV proton and electron beams using simple fluid model for target normal sheath acceleration. Physics of Plasmas, 17(7). https://doi.org/10.1063/1.3459063

Ticoş, D., Scurtu, A., Oane, M., Diplaşu, C., Giubega, G., Călina, I., & Ticoş, C. M. (2019). *Complementary dosimetry for a 6 MeV electron beam.* Results in Physics, 14(February), 102377. https://doi.org/10.1016/j.rinp.2019.102377

Volpe, L., Fedosejevs, R., Gatti, G., Pérez-Hernández, J. A., Méndez, C., Apiñaniz, J., …& Roso, L. (2019). *Generation of high energy laser-driven electron and proton sources with the 200 TW system VEGA 2 at the Centro de Laseres Pulsados.* High Power Laser Science and Engineering, 7, 6–11. https://doi.org/10.1017/hpl.2019.10

Xiao, K. D., Zhou, C. T., Jiang, K., Yang, Y. C., Li, R., Zhang, H., … He, X. T. (2018). *Multidimensional effects on proton acceleration using high-power intense laser pulses*. Physics of Plasmas, 25(2). https://doi.org/10.1063/1.5003619

Zell, H. (2017) National Aeronautics and Space Administration, Page Last Updated: Aug. 4, https://www.nasa.gov/mission_pages/rbsp/science/rbsp-spaceweather.html

Zhou, D., Sullivan, O. D., Semones, E., Zapp, N., Wang, S., Liu, ,,, &Benton, E. R. (2011). *Radiation of Cosmic Rays Measured on the Inter national Space Station 1 Introduction 2 LET Spectr um Method 3 Matroshka Experiments*. 6, 107–110. https://doi.org/10.7529/ICRC2011/V06/1248

Narici, L., Casolino, M., Fino, L. Di, Larosa, M., Picozza, P., Rizzo, A., & Zaconte, V. (2017). *Performances of Kevlar and Polyethylene as radiation shielding on-board the International Space Station in high latitude radiation environment*. Scientific Reports, (March), 1–11. https://doi.org/10.1038/s41598-017-01707-2